\documentclass{moriond}
\usepackage{amsmath,amsbsy,amsfonts,amsmath,latexsym,amssymb}
\usepackage{mathrsfs}

\newenvironment{Eqnarray}%
         {\arraycolsep 0.14em\begin{eqnarray}}{\end{eqnarray}}
\let\Re\relax

\DeclareMathOperator{\Re}{Re}

\def\half{\tfrac12}
\def\eq#1{eq.~(\ref{#1})}

\def\beq{\begin{equation}}
\def\eeq{\end{equation}}
\def\beqa{\begin{Eqnarray}}
\def\eeqa{\end{Eqnarray}}

\def\vev#1{\langle #1 \rangle}
\def\nn{\nonumber}
\def\cbma{c_{\beta-\alpha}}
\def\sbma{s_{\beta-\alpha}}
\def\cba{\cos(\beta-\alpha)}
\def\sba{\sin(\beta-\alpha)}

\def\lsub#1{\ifmath{_{\lower2.5pt\hbox{$\scriptstyle #1$}}}}
\def\phm{\phantom{-}}

\newcommand{\Photo}{\includegraphics[height=35mm]{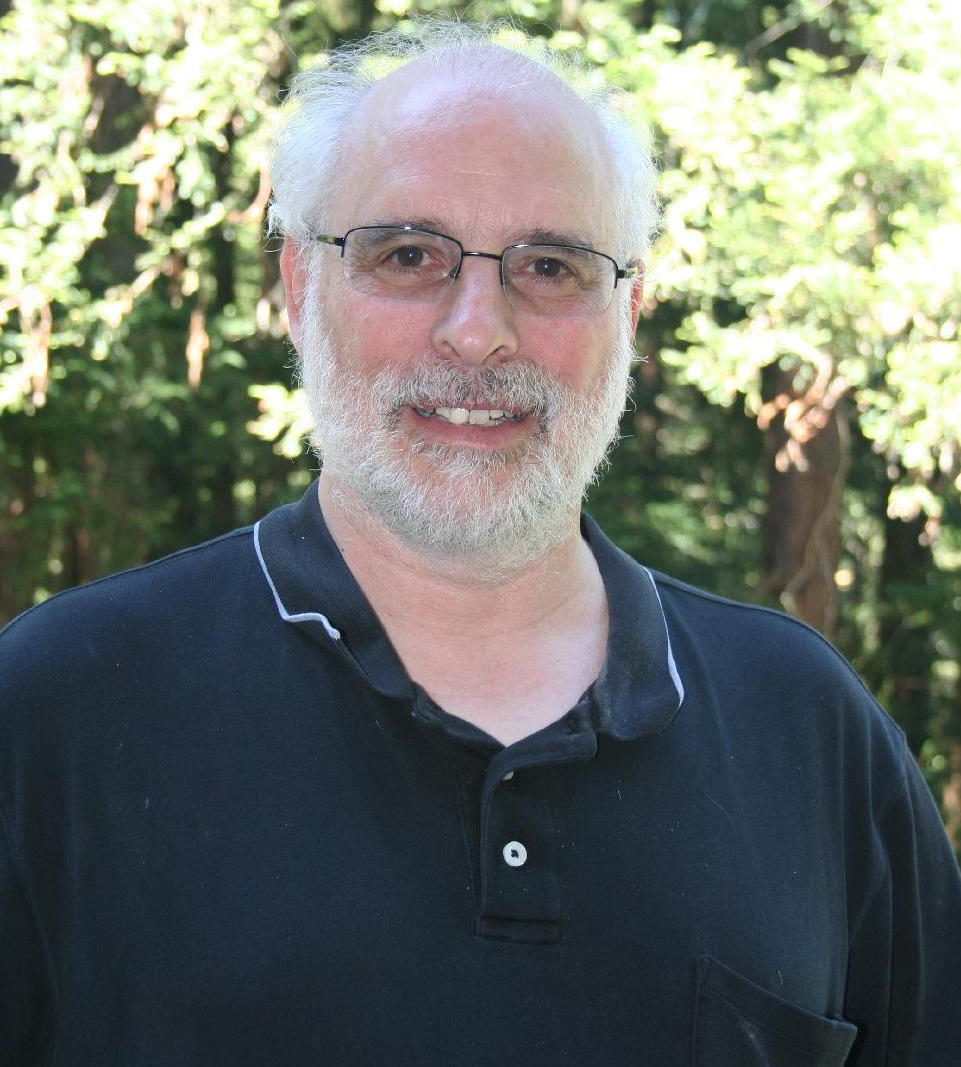}}

\begin{document}
\vspace*{4cm}
\title{APPROXIMATE HIGGS ALIGNMENT WITHOUT DECOUPLING}

\author{ HOWARD E. HABER }

\address{Santa Cruz Institute for Particle Physics, University of California,\\
1156 High Street, Santa Cruz, CA 95064 USA}

\maketitle\abstracts{
The properties of the Higgs boson $h(125)$ observed in LHC data are consistent with
Standard Model predictions.  Consequently, if additional Higgs bosons are present, then the direction of $h(125)$ in field space must be approximately aligned with the direction of the Higgs vacuum expectation value.  One way to achieve approximate Higgs alignment is in the decoupling limit, where all additional Higgs scalars are significantly heavier than the $h(125)$.  In this presentation, the viability of approximate Higgs alignment \textit{without} decoupling is addressed.
}

\section{Introduction}

With the discovery of the Higgs boson in 2012,\cite{Aad:2012tfa,Chatrchyan:2012xdj} it appears that all the fundamental particles that comprise the Standard Model (SM) have been discovered.   
Of course, it is straightforward to add additional Higgs scalars to the SM.  However, an extended Higgs sector is already highly constrained.  For example, the electroweak $\rho$ parameter is very close to 1, which strongly suggests that additional Higgs multiplets consist of hypercharge one doublets and/or singlets under the electroweak gauge group.\cite{Gunion:1989we}  
Moreover, the extended Higgs sector must contain a SM-like $h(125)$,\cite{Khachatryan:2016vau}
whereas evidence for additional scalar states, either via direct production or by virtual exchange, must have (so far) escaped detection in LHC data.  

\section{The alignment limit of the 2HDM}

In this presentation, we shall focus on the two-Higgs doublet model (2HDM) as a prototype for an extended Higgs sector.
The 2HDM consists of two hypercharge one, doublet fields $\Phi_1$ and $\Phi_2$
After minimizing the scalar potential, $\vev{\Phi_i^0}=v_i/\sqrt{2}$ (for $i=1,2$), with
$v\equiv (|v_1|^2+|v_2|^2)^{1/2}=2m_W/g=246$~GeV.  It is convenient to introduce the Higgs basis fields,\cite{Branco:1999fs,Davidson:2005cw}
\beq
H_1=\begin{pmatrix}H_1^+\\ H_1^0\end{pmatrix}\equiv \frac{v_1^* \Phi_1+v_2^*\Phi_2}{v}\,,
\qquad\quad H_2=\begin{pmatrix} H_2^+\\ H_2^0\end{pmatrix}\equiv\frac{-v_2 \Phi_1+v_1\Phi_2}{v}
 \,,
 \eeq
 such that 
  $\vev{H_1^0}=v/\sqrt{2}$ and $\vev{H_2^0}=0$.  
  The Higgs basis is uniquely defined up to an overall rephasing, $H_2\to e^{i\chi} H_2$.  
  In the Higgs basis, the scalar potential is
given by,

\beqa \mathcal{V}&=& Y_1 H_1^\dagger H_1+ Y_2 H_2^\dagger H_2 +[Y_3
H_1^\dagger H_2+{\rm h.c.}]
+\half Z_1(H_1^\dagger H_1)^2+\half Z_2(H_2^\dagger H_2)^2
+Z_3(H_1^\dagger H_1)(H_2^\dagger H_2)
\nn\\
&&\quad 
+Z_4( H_1^\dagger H_2)(H_2^\dagger H_1)+\left\{\half Z_5 (H_1^\dagger H_2)^2 +\big[Z_6 (H_1^\dagger
H_1) +Z_7 (H_2^\dagger H_2)\big] H_1^\dagger H_2+{\rm
h.c.}\right\}\,.\label{higgspot}
\eeqa
The minimization of the scalar potential yields $Y_1=-\half Z_1 v^2$ and $Y_3=-\half Z_6 v^2$.

The neutral scalar field $H_1^0$ is \textit{aligned} in field space with the Higgs vacuum expectation value.  If $\sqrt{2}\,\Re H_1^0-v$ were a mass eigenstate, then its tree-level properties would coincide with the Higgs boson of the SM.
For simplicity, we assume a CP-conserving scalar potential (where the scalar potential parameters are simultaneously real after an appropriate rephasing of $H_2$).
The CP-even Higgs squared-mass matrix is,\cite{Branco:1999fs,Haber:2006ue}
$$
\mathcal{M}_H^2=\begin{pmatrix} Z_1 v^2 & \quad Z_6 v^2 \\  Z_6 v^2 & \quad m_A^2+Z_5 v^2\end{pmatrix}.
$$
where $m_A$ is the mass of the CP-odd Higgs scalar.

The CP-even Higgs bosons are $h$ and $H$ with $m_h\leq m_H$.
Approximate alignment arises in two limiting cases.\cite{Craig:2013hca,Haber:2013mia,Carena:2013ooa}
\underline{Case 1}: $m^2_A\gg (Z_1-Z_5)v^2$.  This is the \textit{decoupling limit},\cite{Gunion:2002zf} where $h$ is SM-like and
$m^2_A\sim m^2_H\sim m^2_{H^\pm}\gg m^2_h\simeq Z_1 v^2$. \underline{Case 2}:
$|Z_6|\ll 1$.  Then, $h$ is SM-like if $m_A^2+(Z_5-Z_1)v^2>0$; otherwise, $H$ is SM-like.
This is alignment with or without decoupling, depending on how large $m_A$ is.  
In particular, the CP-even neutral scalar mass eigenstates are,
\beq
\begin{pmatrix} H\\ h\end{pmatrix}=\begin{pmatrix} \cbma & \,\,\, -\sbma \\
\sbma & \,\,\,\phantom{-}\cbma\end{pmatrix}\,\begin{pmatrix} \sqrt{2}\,\,\Re H_1^0-v \\ 
\sqrt{2}\,\Re H_2^0
\end{pmatrix}\,,
\eeq
where $\cbma\equiv\cba$ and $\sbma\equiv\sba$ are defined in terms of the mixing angle $\alpha$ that diagonalizes the CP-even Higgs squared-mass matrix when expressed in the $\Phi_1$--$\Phi_2$ basis of scalar fields, $\{\sqrt{2}\, \Re\Phi_1^0-v_1\,,\,\sqrt{2}\,\Re\Phi_2^0-v_2\}$, and $\tan\beta\equiv v_2/v_1$.
Since the SM-like Higgs boson must be approximately $\sqrt{2}\,\Re H_1^0-v$, it follows that
$h$ is SM-like if $|\cbma|\ll 1$ (corresponding to alignment with or without 
decoupling, depending on how large $m_A$ is).  Likewise, $H$ is SM-like if $|\sbma|\ll 1$ (which is only possible in the case of alignment without decoupling).

Exact tree-level Higgs alignment corresponds to $Z_6=0$, which is satisfied in the inert doublet model (IDM).\cite{Barbieri:2006dq}  This is a special case of the 2HDM in which there is 
an unbroken  $\mathbb{Z}_2$ symmetry in the Higgs basis such that $H_2$ is the only $\mathbb{Z}_2$-odd field.  
In the IDM, the SM Higgs boson resides entirely in~$H_1$.
In contrast, approximate tree-level alignment without decoupling in a generic 2HDM would appear to depend on a judicious choice of
model parameters.   However, a more satisfying scenario would be one in which the latter is achieved as a consequence of an approximate symmetry.  Thus, we now explore the possible symmetries of the 2HDM scalar potential that can lead naturally to tree-level Higgs alignment, and the symmetry-breaking mechanisms that can maintain 
approximate alignment, consistent with the LHC Higgs data.

\section{A symmetry origin for Higgs alignment}

One can reduce the number of 2HDM parameters by imposing additional symmetries on the scalar potential in the $\Phi_1$--$\Phi_2$ basis.\cite{Ivanov:2007de,Ferreira:2009wh}  In this basis, the parameters of the scalar potential are three squared-mass  parameters, $m_{11}^2$,
$m_{22}^2$ and $m_{12}^2$ (the analogs of the $Y_i$ of \eq{higgspot}), and seven quartic couplings, $\lambda_i$, $i=1,2,\ldots,7$ (the analogs of the $Z_i$ of \eq{higgspot}).  The possible Higgs family symmetries are:
(i) $\mathbb{Z}_2$: $\Phi_1\to \Phi_1$ and $\Phi_2\to -\Phi_2$; (ii) $\Pi_2$: $\Phi_1\longleftrightarrow\Phi_2$;
(iii) U(1)$_{\rm PQ}$ [Peccei-Quinn~\!\cite{Peccei:1977ur}]: $\Phi_1\to e^{-i\theta}\Phi_1$ and $\Phi_2\to e^{i\theta}\Phi_2$; (iv) SO(3): $\Phi_i\to U_{ij}\Phi_j$, for $U\in$~U(2)/U(1)$_{\rm Y}$ (the scalar potential is automatically invariant with respect to the hypercharge U(1)$_{\rm Y}$).  In addition, one can also consider generalized CP symmetries: (i) CP1: $\Phi_i\to\Phi_i^*$,
(ii) CP2: $\Phi_1\to \Phi_2^*$ and $\Phi_2\to -\Phi_1^*$; (iii) CP3: $\Phi_1\to \Phi_1^*\cos\theta+\Phi_2^*\sin\theta$ and $\Phi_2\to -\Phi_1^*\sin\theta+\Phi_2^*\cos\theta$ (for $0<\theta<\half\pi$).  The constraints of each of these symmetries on the scalar potential parameters in the $\Phi_1$--$\Phi_2$ basis are easily obtained (see Table 1 of Ref.~13).
Moreover, applying $\mathbb{Z}_2$ and $\Pi_2$ simultaneously is equivalent to a CP2-symmetric potential in a different basis.  Similarly, applying U(1)$_{\rm PQ}$ and $\Pi_2$ simultaneously is equivalent to a CP3-symmetric potential in a different basis.\cite{Ferreira:2009wh}

The parameters of the CP-conserving scalar potential in
the $\Phi_1$--$\Phi_2$ basis are related to the corresponding Higgs basis parameters; e.g.,\cite{Haber:2015pua}
\beqa
Y_3 &=& \half (m_{22}^2-m_{11}^2) s_{2\beta}-m_{12}^2 c_{2\beta}\,, \\
Z_6&=&-\half\bigl[\lambda_1 c_\beta^2-\lambda_2 s_\beta^2-\lambda_{345} c_{2\beta}\bigr]s_{2\beta}+\lambda_6 c_\beta c_{3\beta}+\lambda_7 s_\beta s_{3\beta}\,,\label{zeesix}
\eeqa
where $\lambda_{345}\equiv\lambda_3+\lambda_4+\lambda_5$, $s_\beta\equiv\sin\beta$, $c_\beta\equiv\cos\beta$, etc.
Due to the scalar potential minimum condition, exact alignment (i.e., $Z_6=0$) implies that $Y_3=0$.  The latter can be achieved if the scalar potential exhibits a CP2, CP3 or SO(3) symmetry, in which case  $m_{11}^2=m_{22}^2$ and $m_{12}^2=0$.
In the case of CP2, setting $Z_6=0$ in \eq{zeesix} determines the value of $\tan\beta$.  In the CP3 and SO(3) cases, $\lambda_1=\lambda_2=\lambda_{345}$ and $\lambda_6=\lambda_7=0$.  Then,
$Z_6=0$ is satisfied independently of $\tan\beta$, corresponding to the condition of natural alignment introduced by Bhupal Dev and Pilaftsis.\cite{Dev:2014yca}

In order to specify a complete model, one must also exhibit the Higgs--fermion Yukawa couplings.  In the case of the IDM,
the fermions are taken to be $\mathbb{Z}_2$-even states, which do not couple to the Higgs basis field $H_2$.
In contrast, the extension of the CP2, CP3 or SO(3) symmetries to the Yukawa interactions is problematic, resulting in a massless quark or some other phenomenologically untenable feature.\cite{Ferreira:2010bm}

\section{A model of approximate Higgs alignment}

Consider the 2HDM with a CP2-symmetric scalar potential, which can be realized in another basis as 
a $\mathbb{Z}_2\otimes\Pi_2$ discrete symmetry, where
$m_{11}^2=m_{22}^2$, $\lambda_1=\lambda_2$,  $m_{12}^2=\lambda_6=\lambda_7=0$, and $\lambda_5$ is real.
To extend this symmetry to the Yukawa sector, we introduce mirror fermions.\cite{Draper:2016cag}
The SM fermions are denoted by lower case letters (e.g.~left-handed doublet fields $q$ and right-handed singlet fields $u$ and~$d$), and mirror fermions by upper case letters.  
Under the discrete symmetries,
\beqa
\Pi_2:&&  \Phi_1\longleftrightarrow\Phi_2\,,\quad q \longleftrightarrow q, \quad u \longleftrightarrow U,\phm \nonumber\quad \overline{U}\longleftrightarrow \overline{U}\,, \\
\mathbb{Z}_2:&&  \Phi_1\to\Phi_1\,,\quad \Phi_2\to -\Phi_2\,,\quad q \to q,  \quad u \to -u, \quad U \to U,\quad \overline{U}\to \overline{U}\,, \nonumber
\eeqa
where $\overline{U}$ is in the representation conjugate to $U$ (to avoid anomalies).
The Yukawa couplings consistent with the $\mathbb{Z}_2\otimes\Pi_2$ discrete symmetry are,\footnote{The down-type quarks and leptons can also be included by introducing the appropriate mirror fermions.}
\beq
\mathscr{L}_{\rm Yuk} \, \supset \, y_t \left(q \Phi_2 u + q \Phi_1 U \right) + \rm{h.c.}
\eeq
In addition, we introduce an explicit mass term,
$
\mathscr{L}_{\rm mass}= M_U U\overline{U}+{\rm h.c.}
$
(with $M_U$ large enough to be consistent with the LHC experimental limits on mirror fermion masses),
which preserves the $\mathbb{Z}_2$ but explicitly breaks the $\Pi_2$ discrete symmetry.  This symmetry breaking is soft, so that 
$m_{22}^2-m_{11}^2$ is protected from quadratic sensitivity to the cutoff scale $\Lambda$,
$$
\Delta m^2 \equiv m_{22}^2-m_{11}^2\sim -\frac{3y_t^2M_U^2}{4\pi^2}\ln(\Lambda/M_U)\;,
$$
neglecting finite threshold corrections proportional to $M_U^2$.  Due to the unbroken $\mathbb{Z}_2$ symmetry, a nonzero $m_{12}^2$ is not generated in this approximation. Integrating out the mirror fermions below the scale $M_U$, one also generates a splitting between $\lambda_1$ and $\lambda_2$.  However, this is a small correction, which has a negligible impact on our analysis.

The important parameters of the scalar potential are:
$m^2\equiv \half(m_{11}^2+m_{22}^2)$, $\Delta m^2\equiv m_{22}^2-m_{11}^2$ and
$R \equiv \lambda_{345}/\lambda$ (where $\lambda\equiv\lambda_1=\lambda_2$).
We impose $\lambda>0$ and $R>-1$ to ensure that the vacuum is bounded from below.
Solving for the scalar potential minimum, there are two possible phases:
(i)
the inert phase, where the $\mathbb{Z}_2$ is unbroken, corresponding to the IDM; and 
(ii) a mixed phase where both $v_1\neq 0$ and $v_2\neq 0$.
In the case of the mixed phase, $m^2=-\tfrac{1}{4}\lambda(1+R)v^2$ and
\beq
\tan\beta\equiv \frac{v_2}{v_1}=\sqrt{\frac{1-\epsilon}{1+\epsilon}}\,,\qquad \text{where\quad
$\epsilon=\cos 2\beta=\frac{2\Delta m^2}{\lambda(1-R)v^2}$}\,.
\eeq
The positivity of $v_{1}^2$ and $v_{2}^2$ and the curvature at the extremum requires $|R|<1$ and 
$|\epsilon|<1$.

The Higgs boson mass spectrum of the mixed phase is,
\beq
m_{h,H}^2=\tfrac{1}{2} \lambda v^2\bigl[1\mp\sqrt{R^2+(1-R^2)\epsilon^2}\,\bigr]\,,\qquad
m_A^2= |\lambda_5| v^2\,,\qquad 
m_{H^\pm}^2=m_A^2-\half(\lambda_4-\lambda_5)v^2\,.
\eeq
If $h$ is SM-like, then $-1<R<-\epsilon^2/(1-\epsilon^2)$ and
\beq
c_{\beta-\alpha}\simeq \frac{\epsilon(1-R)}{2|R|}+\mathcal{O}(\epsilon^2)\,.
\eeq
In particular, the alignment limit favors small $|\epsilon|$, which yields $\tan\beta\sim\mathcal{O}(1)$.
In this parameter regime,
\beq
m_H^2\simeq m_h^2\left(\frac{1+|R|}{1-|R|}+\mathcal{O}(\epsilon^2)\right)\,,
\eeq
corresponding to approximate alignment without decoupling as long as $|R|$ is not too close to $1$.

One can also construct a model of approximate alignment without decoupling by employing a softly broken CP3 symmetry, augmented with the corresponding mirror fermions.\cite{prep} Further details on the phenomenology of the CP2 and CP3 models can be found in Refs.~19 and~20.   

\section*{Acknowledgments}

This presentation is based on works in collaboration with Patrick Draper, Pedro Ferreira, Jo\~ao Silva, and Joshua Ruderman.
H.E.H. is supported in part by the U.S. Department of Energy grant
number DE-SC0010107.

\end{document}



